\documentclass[9pt,conference]{IEEEtran}

\usepackage{waspaa25} 

\usepackage{comment}
\usepackage{graphicx}
\usepackage{subcaption}
\usepackage[T1]{fontenc}
\usepackage{bm} 

\title{Diffused Responsibility: Analyzing the Energy Consumption of Generative Text-to-Audio Diffusion Models}

\name{Riccardo Passoni$^{1}$,
      Francesca Ronchini$^{1}$,
      Luca Comanducci$^{1}$,
      Romain Serizel$^{2}$,
      Fabio Antonacci$^{1}$}
\address{
$^{1}$Dipartimento di Elettronica, Informazione e Bioingegneria - Politecnico di Milano, Milan, Italy\\
$^{2}$Université de Lorraine, CNRS, Inria, Loria, Nancy, France
}

\begin{document}

\maketitle

\begin{abstract}
Text-to-audio models have recently emerged as a powerful technology for generating sound from textual descriptions. However, their high computational demands raise concerns about energy consumption and environmental impact. In this paper, we conduct an analysis of the energy usage of 7 state-of-the-art text-to-audio diffusion-based generative models, evaluating to what extent variations in generation parameters affect energy consumption at inference time. We also aim to identify an optimal balance between audio quality and energy consumption by considering Pareto-optimal solutions across all selected models. Our findings provide insights into the trade-offs between performance and environmental impact, contributing to the development of more efficient generative audio models.
\end{abstract}

\section{Introduction}
\label{sec:intro}

Audio generative models have led to remarkable progress across numerous audio processing tasks~\cite{purwins2019deep}. A recent development in this domain is the introduction of Text-to-Audio (TTA) models, which can synthesize audio content from descriptive text prompts~\cite{dong2024generative, ma2024foundation}. While these models have achieved impressive results, they also require significant computational demands~\cite{douwes2021energy, douwes2025energy, dong2024generative}. A key concern is the high energy consumption required to train and deploy such models, which contributes to an increasing carbon footprint~\cite{thompson2020computational, schwartz2020green, holzapfel2024green}.
Until very recently, the environmental impact of deep learning models has largely been dominated by the persistent demand for high accuracy and effectiveness~\cite{douwes2023quality}. In the last years, there has been a rise of concern about the environmental impact of deep learning within audio signal processing communities. The relation between energy consumption and the performance of data-driven sound event detection models has been explored in several studies~\cite{serizel2023performance, douwes2023quality, parcollet2021energy, gabrielli2024sustainability, douwes2023monitoring, ronchini2024performance, douwes2025energy}. Douwes et al. investigated the energy implications of neural audio synthesis models~\cite{douwes2023quality}, while concerns around sustainability and energy use have been raised in natural language processing~\cite{parcollet2021energy} and the Internet of Sounds~\cite{gabrielli2024sustainability} communities. Similarly, Holzapfel et al. provide an initial estimate of the energy required to train deep learning models for music information retrieval tasks~\cite{holzapfel2024green}. 

Although TTA generative models have rapidly advanced in recent years, particularly in terms of generative quality, the energy consumption required for their training and deployment has received little attention. 
To our knowledge, only related work on multimodal image generation has begun to address this issue~\cite{luccioni2024power}. Motivated by this gap, we propose an analysis of the energy consumption of $7$ state-of-the-art diffusion-based~\cite{yang2023diffusion} TTA models to better understand their environmental implications, focusing specifically on the inference phase. While most existing related research on data-driven models has concentrated on the training phase due to its high energy demands, inference can have an equal or even greater impact. In fact, although a single inference requires limited computation compared to training, it is performed far more frequently, particularly in real-world user-oriented applications (e.g., interactions with ChatGPT), thus leading to significant cumulative energy consumption~\cite{luccioni2024power, douwes2023environmental}. To investigate this, we conducted two experiments analyzing how the batch size and the number of inference steps relate to the energy consumption. Recognizing that efficiency and audio quality are both essential, especially for practical deployment, we then explored the trade-off between these objectives by analyzing the Pareto frontier of the evaluated TTA models, following the approach adopted in~\cite{douwes2023quality}. This method enables the identification of optimal configurations in which enhancing one objective (e.g., audio quality) may require sacrificing another (e.g., energy efficiency). These insights aim to support a more sustainable use of TTA models and help users make informed, application-specific choices when selecting models for audio generation.
The code used is available at 
\href{https://github.com/rickgiantsteps/audiodiffusion-energy-consumption}{https://github.com/rickgiantsteps/audiodiffusion-energy-consumption}.

\section{METHODOLOGY AND EXPERIMENTAL SETUP}
\label{subsec:exp}

This section outlines the methodology and experimental setup, and presents the Pareto optimization analysis. Section~\ref{subsec:TTAsel} offers a brief overview of the TTA models included in this study. Section~\ref{subsec:eva} presents the framework used to evaluate energy consumption, while Section~\ref{subsec:pa} outlines the methodology for analyzing the Pareto frontier.

\subsection{TTA models selected}
\label{subsec:TTAsel}

We analyzed 7 state-of-the-art models based on the diffusion approach to generate audio: AudioLDM, AudioLDM2, Make-an-Audio, Make-an-Audio-2, Stable Audio Open, Tango, and Tango2. AudioLDM ~\cite{audioldm} is built on a latent space that learns continuous audio representation from Contrastive Language-Audio Pre-trained (CLAP) embeddings~\cite{elizalde2023clap}. The model is improved in AudioLDM2~\cite{audioldm2}, where a universal audio representation is used to provide a robust foundation for the audio generation task. Make-an-Audio~\cite{maa} introduces a pseudo-prompt enhancement strategy to generate prompts better aligned with audio. It also employs a spectrogram autoencoder to predict self-supervised representations instead of raw waveforms, enabling efficient compression and high-level semantic understanding of complex audio signals. Built upon this foundation, Make-an-Audio-2~\cite{maa2} further enhances semantic alignment and temporal consistency. Tango~\cite{tango} adopts the variational autoencoder from AudioLDM and replaces the CLAP model with FLAN-T5~\cite{longpre2023flan}, a fine-tuned large language model. This design aims to match or surpass the performance of previous models while requiring significantly less training data. Tango2 places explicit emphasis on capturing the presence of concepts or events and preserving their temporal order in the generated audio relative to the input prompt. Stable Audio Open~\cite{evans2025stable} incorporates an autoencoder that reduces waveforms to a more compact sequence, a T5-based text embedding for conditioning on text, and a transformer-driven diffusion model that functions within the autoencoder's latent space. Table~\ref{tab:table1} shows the selected checkpoint (when multiple are available), the number of model parameters, and the inference speed (on the GPU used in this study) for 10-second audio generation.

\begin{table}[t!]
\caption{TTA model checkpoints, number of parameters, and inference speed (considering a $10$-second audio clip using $100$ denoising steps)}
\label{tab:table1}
\centering
\sisetup{
    table-number-alignment = center,
    round-mode = places,
    round-precision = 2,
}
\begin{tabular}{cccc}
\toprule
\textbf{Model} & \textbf{Checkpoint} & \textbf{Param.} & \textbf{Inference Speed (s)} \\
\midrule
AudioLDM & audioldm-s-full & 421M & $2.82$  \\
AudioLDM2 & audioldm2-full & 1.1B & $6.06$ \\
Make-an-Audio & - & 453M & $2.63$ \\
Make-an-Audio-2 & - & 937M & $2.62$  \\
Stable Audio Open & - & 1.21B & $8.16$ \\
Tango & tango-full & 1.2B & $12.06$ \\
Tango2 & tango2-full & 1.2B & $12.09$ \\
\bottomrule
\end{tabular}
\end{table}

\subsection{Energy Consumption evaluation}
\label{subsec:eva}
In recent years, different tools and packages have been introduced to evaluate energy-related metrics of deep learning models~\cite{anthony2020carbontracker, rasley2020deepspeed, schmidt2021codecarbon, zhu2019thop}. A detailed review of these tools is beyond the scope of this paper. However, readers can refer to~\cite{douwes2023quality, douwes2023monitoring} for an overview of the available options. Given our focus on tracking the energy consumption of TTA models, we employ the CodeCarbon toolkit~\cite{benoit_courty_2024_11171501}. The tool estimates energy consumption by monitoring the power usage of GPU, CPU, and RAM and aggregates these to compute the total energy consumed by the computing infrastructure, whether local or cloud-based. In line with previous studies~\cite{douwes2024computation, douwes2025energy}, and given that GPU usage accounts for the majority of energy consumption in deep learning generative models, 
we report only GPU energy usage for all TTA models. Energy values are presented in kilowatt-hours (kWh).

\subsection{Pareto frontier}
\label{subsec:pa}
Audio quality is a critical factor in the evaluation of TTA models. To capture the trade-offs between audio quality and energy consumption, we apply Pareto optimization as a means of assessing the performance of TTA models. Pareto optimization is a well-established method for addressing multi-objective optimization problems~\cite{douwes2023quality, douwes2021energy}. 
A configuration is considered Pareto efficient if any further improvement in one objective would result in a degradation of the other. The Pareto frontier shows all these best trade-offs, allowing us to find the setups that balance audio quality and energy use most effectively. The reader is referred to~\cite{douwes2023quality} for further details on the Pareto frontier. To perform Pareto optimization of the models, we generated two datasets using the TTA models, computed the CLAP score~\cite{elizalde2023clap} and Fréchet Audio Distance (FAD)~\cite{kilgour2018fr} for the generated data, and analyzed how these metrics relate to energy consumption.


\textbf{Datasets.} We used AudioCaps-test and Clotho-eval subsets from AudioCaps~\cite{kim2019audiocaps} and Clotho~\cite{drossos2020clotho} datasets, respectively. AudioCaps is composed of around $46\mathrm{K}$ $10$-second audio clips from AudioSet~\cite{gemmeke2017audio}, each paired with a human-written caption, with the test set containing $4895$ clip/caption pairs. Clotho consists of $4981$ audio clips lasting $15$ to $30$ seconds, each annotated with five captions, and its evaluation set includes $1045$ samples. Both datasets are designed for audio captioning tasks, with captions crowdsourced via Amazon Mechanical Turk from annotators in English-speaking countries.


\textbf{Audio Quality metrics.} We evaluated the audio quality of the TTA models using two widely adopted metrics: CLAP score~\cite{elizalde2023clap} and FAD~\cite{kilgour2018fr}. The CLAP score evaluates the alignment between an audio clip and a textual description using CLAP~\cite{elizalde2023clap}, a model that maps both audio and language into a shared embedding space. FAD is a reference-free metric adapted from the Frechet Inception Distance (FID) used in image generation\cite{heusel2017gans}, and was introduced for music evaluation in~\cite{roblek2019fr}. It measures the distributional distance between generated and real audio using embeddings from a pretrained audio classifier. Low FAD values indicate that the generated audio closely matches real audio samples, while high CLAP scores reflect better semantic alignment between the audio and its corresponding prompt. In our experiments, FAD was calculated using the \textit{fadtk}~\cite{fadtk} framework with \textit{clap-laion-audio} for audio embeddings, while CLAP scores were obtained through the \textit{stable-audio-metrics} toolkit~\cite{stableaudiometrics}. 


\section{Experiments and results}
\label{sec:res}

This section presents the experiments conducted in this study and their results. Each experiment was repeated $5$ times to compute the average and standard deviation of energy consumption. To ensure fair comparisons across models, measurements were limited to the audio generation step, excluding model loading and audio file storage. All experiments were conducted on a single NVIDIA A40 GPU.

\begin{figure*}[t!]
  \centering
  
  \vspace{-2em}  \centerline{\includegraphics[width=\textwidth]{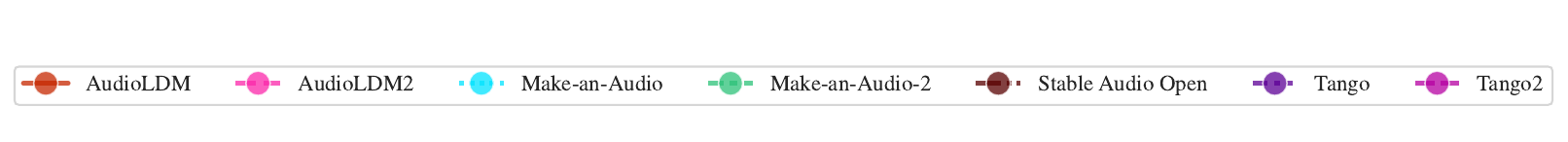}}
  \vspace{-1em}
  \begin{subfigure}[b]{0.495\textwidth}
    \includegraphics[width=\columnwidth]{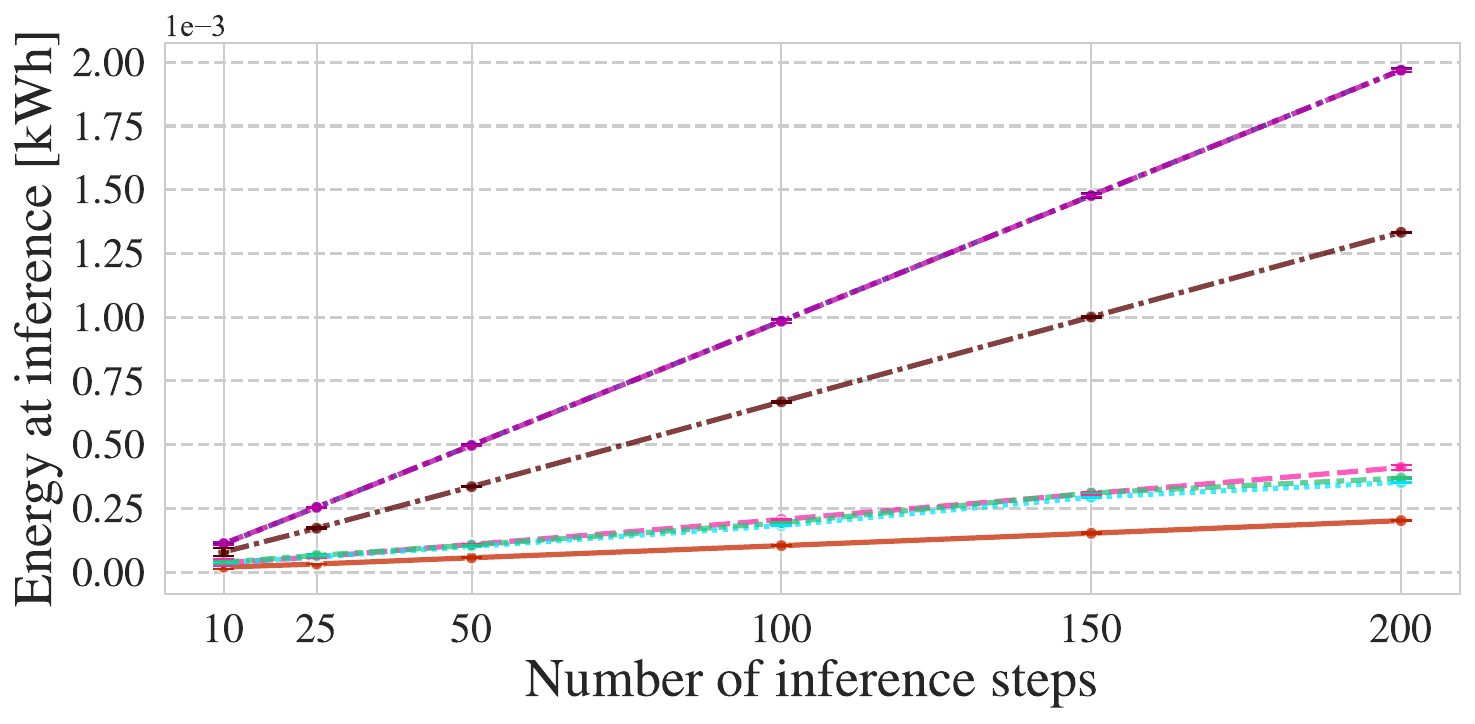}
    \caption{Models with inference steps}
    \label{fig:exp1}
  \end{subfigure}
  \hfill 
  \begin{subfigure}[b]{0.495\textwidth}
    \includegraphics[width=\columnwidth]{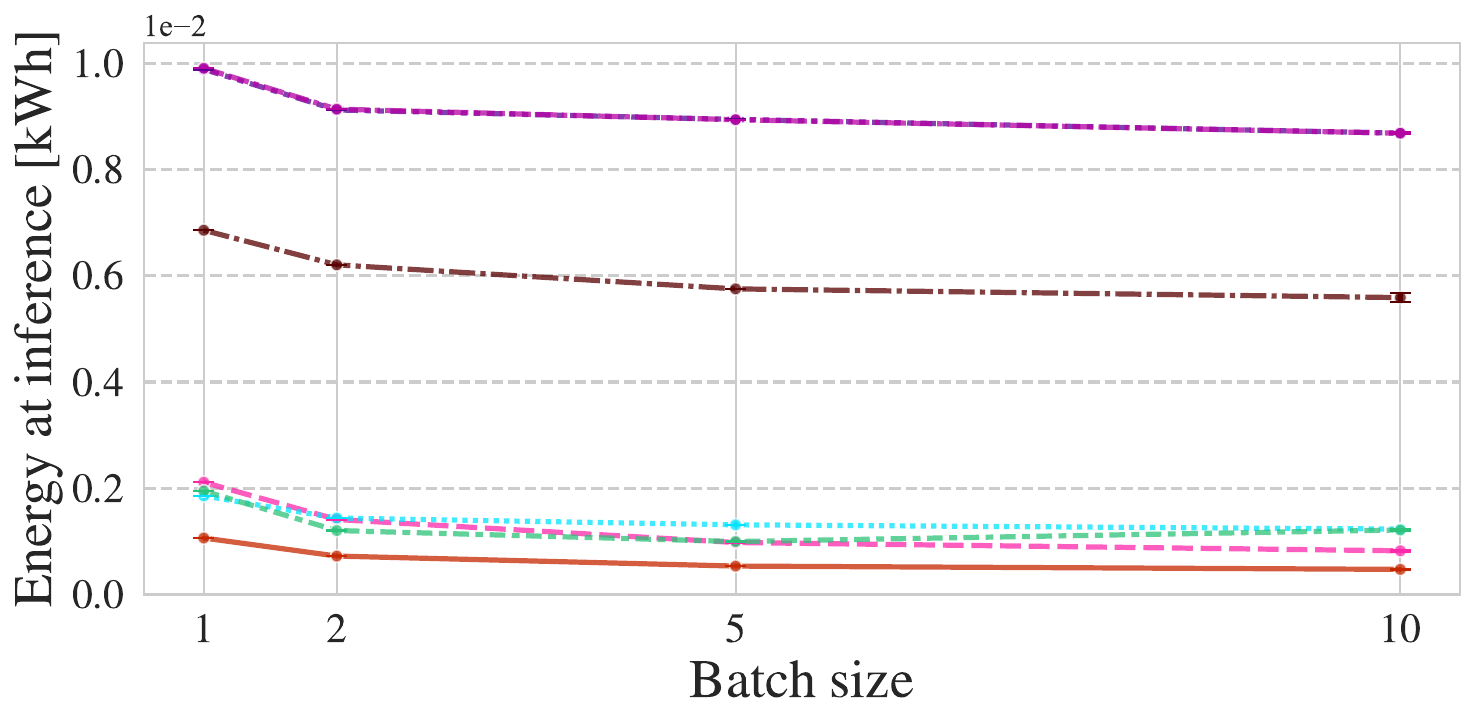}
    \caption{Batch size generation.}
    \label{fig:exp2}
  \end{subfigure}
  \caption{Energy consumption at inference as a function of the number of inference steps (a) and batch size (b). The legend indicates the different TTA models compared in the analysis, with colors representing each model.}  
  \label{fig:exps12}
\end{figure*}

\subsection{Does energy consumption scale with inference steps?}
\label{subsec:exp1}

Many studies on TTA models emphasize the importance of tuning different parameters to improve performance. One of the most important factors for diffusion models is the number of inference steps, which controls how many iterations the model runs during audio generation. While increasing the number of steps can initially improve quality, the benefits might decrease after a certain point~\cite{audioldm}. More steps also lead to longer computation times and higher energy use, without always improving performance. Since this parameter is commonly adjusted to boost quality, we believe it is important to understand how it affects the model’s energy consumption. To investigate this, we generated a separate $10$-second audio sample for each inference step using the same prompt, varying the number of steps across $10$, $25$, $50$, $100$, $150$, and $200$.

Figure~\ref{fig:exp1} presents the results of our first experiment, showing a clear linear relationship between the number of inference steps and energy consumption across all models. Although this trend is consistent, the rate at which energy consumption increases with the number of steps varies depending on each model’s design. Among the models evaluated, AudioLDM stands out as the most energy-efficient, making it particularly well-suited for deployment on energy-sensitive platforms. In contrast, Tango and Tango2 are more energy-intensive and may require further optimization for use in low-power environments. The Make-an-Audio family shows moderate energy consumption, particularly in scenarios involving a high number of inference steps. Interestingly, newer model generations, such as Make-an-Audio2 and AudioLDM2, tend to consume more energy than their predecessors, suggesting a continued prioritization of performance during the development phase, often at the cost of efficiency. Differently, Tango and Tango2 consume the same energy, likely due to their shared architecture, with Tango2 being a fine-tuned version of Tango using diffusion direct preference optimization.

\subsection{Does the batch size affect energy consumption at inference?}
\label{subsec:exp2} 

In certain applications, such as artistic sound generation or creative workflows~\cite{ronchini2024paguri, kamath2024sound}, generating multiple audio samples from the same prompt 
can be useful. This approach enables a broader exploration of the generation space and offers a variety of options. However, it also raises an important question: is it more efficient to generate all samples in a single batch, or should they be produced one at a time? Although this scenario is highly implementation- and hardware-dependent, it remains an important use case to analyze. Understanding whether generating multiple outputs in fewer runs provides any advantages is key to optimizing resource usage during the diffusion process. In this experiment, we set the number of denoising inference steps to $100$ for each generation.
We generated a total of $10$ audio samples for each batch size, testing different batch sizes ranging from $1$ to $10$. Each batch size determines how many samples are produced in a single generation run. 

Figure~\ref{fig:exp2} presents the results of this experiment. As expected, increasing the batch size reduces energy consumption per generated sample, with the most significant drop occurring between batch sizes $1$ and $2$. Beyond this point, particularly after batch size $5$, efficiency gains decrease, and most models show only minor improvements or even slight increases in energy use. This suggests larger batch sizes are less effective in reducing energy consumption, possibly due to parallelization costs. This requires further investigation. These findings highlight the benefits of batching multiple generations together, while also indicating the existence of an optimal batch size for energy efficiency.
In terms of model performance, the results align with trends observed in the previous experiment. AudioLDM remains the most energy-efficient model across all batch sizes, showing excellent scalability and consistently low energy consumption as batch size increases. In contrast, Tango and Tango2 are the most energy-intensive models and the ones that benefit the least from batch processing. Stable Audio Open also consumes significantly more energy than both AudioLDM and the Make-an-Audio family of models. 
\begin{figure*}[t!]
  \centering
  \begin{subfigure}[b]{0.495\textwidth}
    \includegraphics[width=\linewidth]{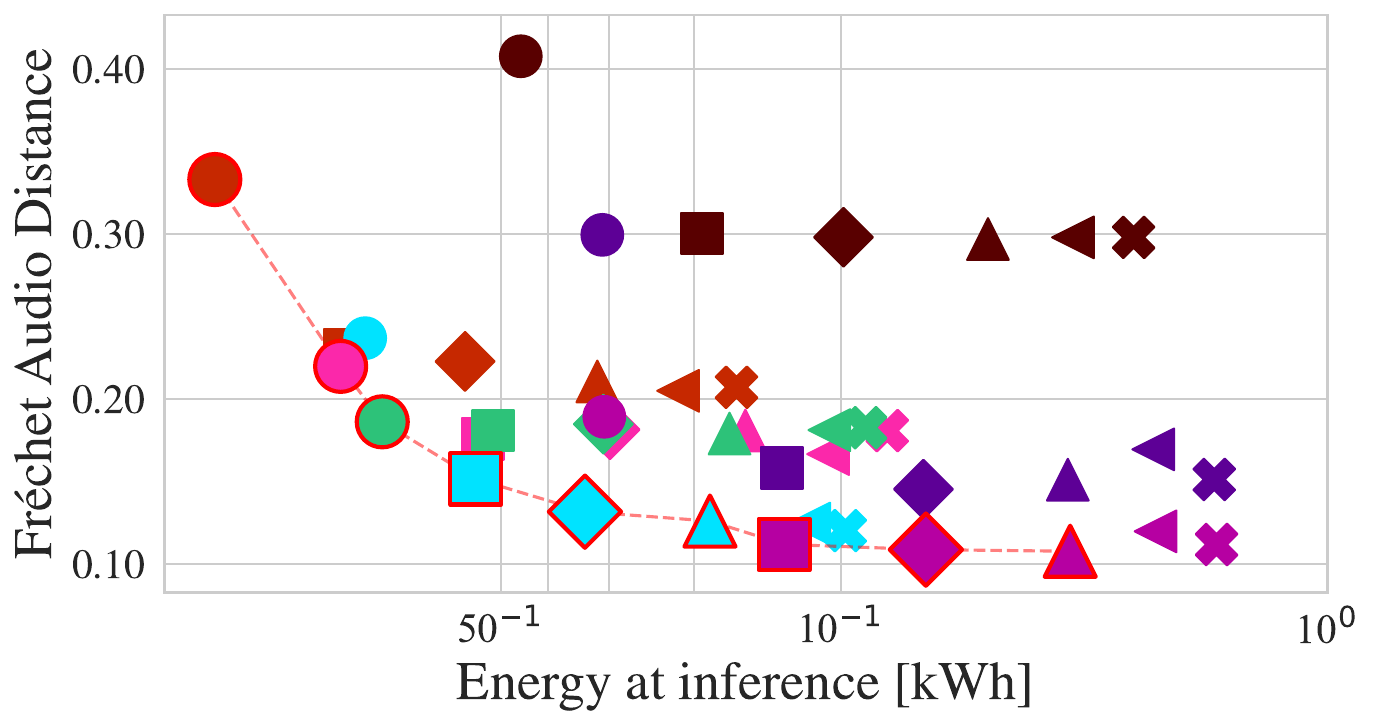}
    \caption{Pareto frontier for Audiocaps based on FAD}
    \label{fig:audiocaps_fad}
  \end{subfigure}
  \hfill
  \begin{subfigure}[b]{0.495\textwidth}
    \includegraphics[width=\linewidth]    {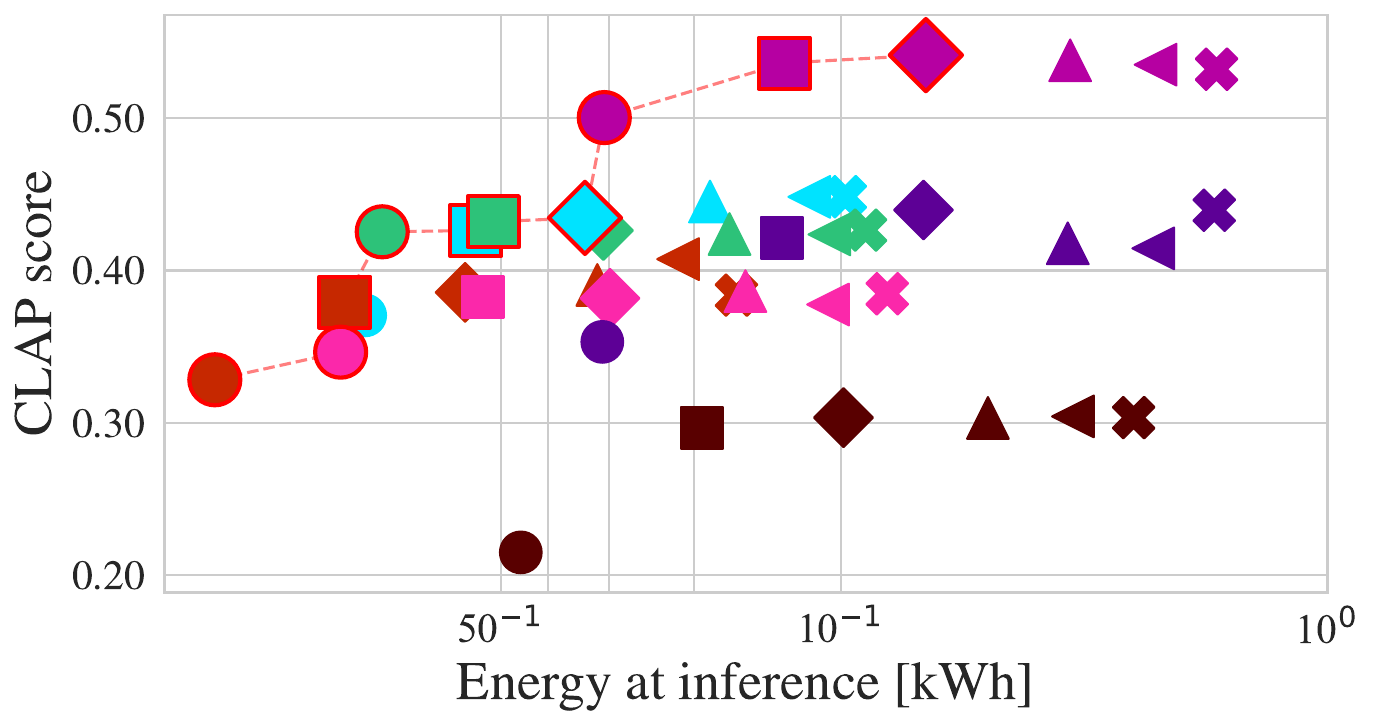}
    \caption{Pareto frontier for Audiocaps based on CLAP}
    \label{fig:audiocaps_clap}
  \end{subfigure}

  \vspace{0.5em}

  \begin{subfigure}[b]{0.495\textwidth}
    \includegraphics[width=\linewidth]{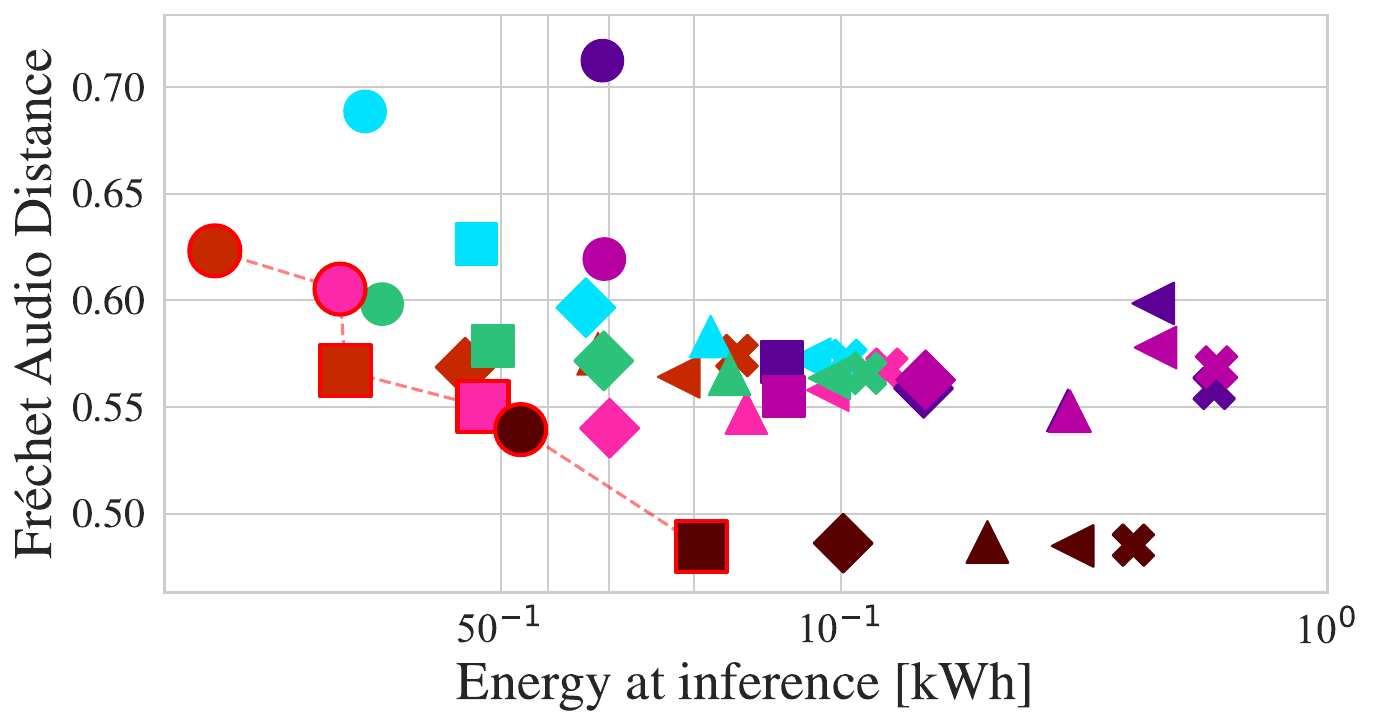}
    \caption{Pareto frontier for Clotho based on FAD}
    \label{fig:clotho_fad}
  \end{subfigure}
  \hfill
  \begin{subfigure}[b]{0.495\textwidth}
    \includegraphics[width=\linewidth]
    {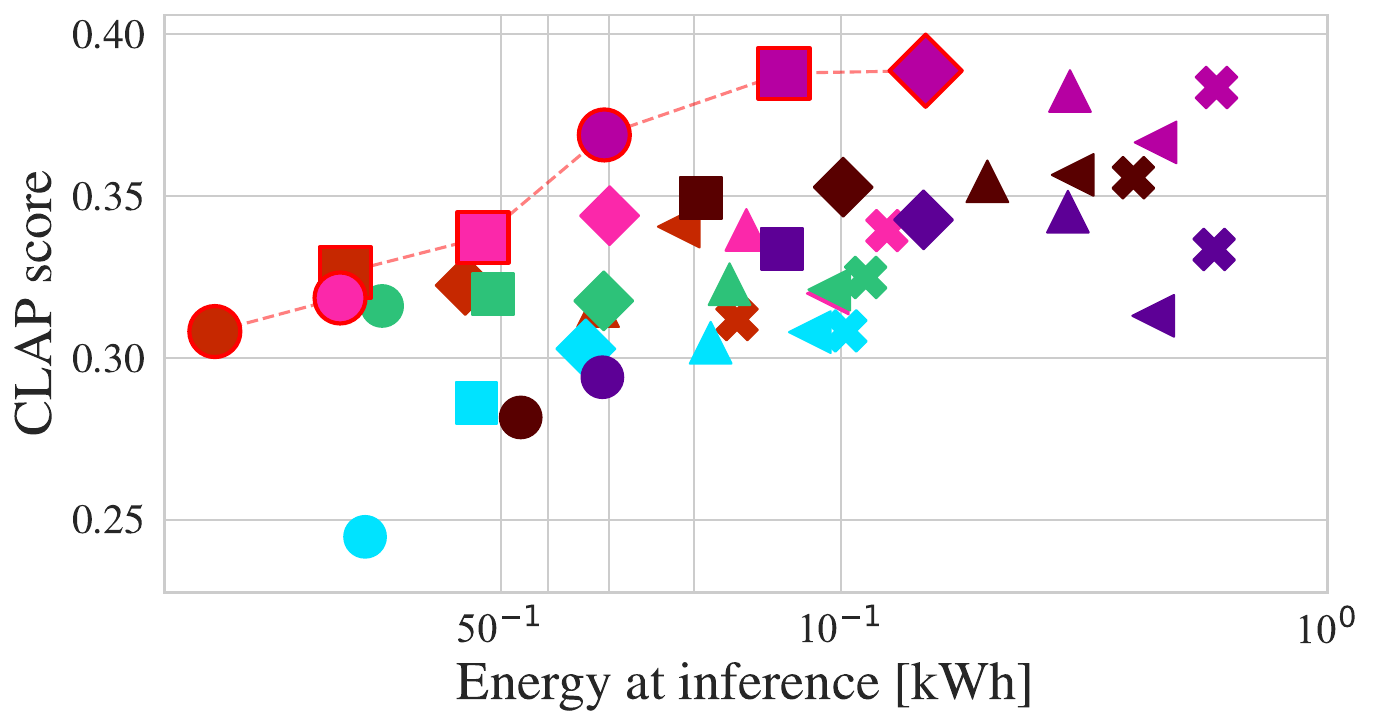}
    \caption{Pareto frontier for Clotho based on CLAP}
    \label{fig:clotho_clap}
  \end{subfigure}


  \includegraphics[width=\textwidth]{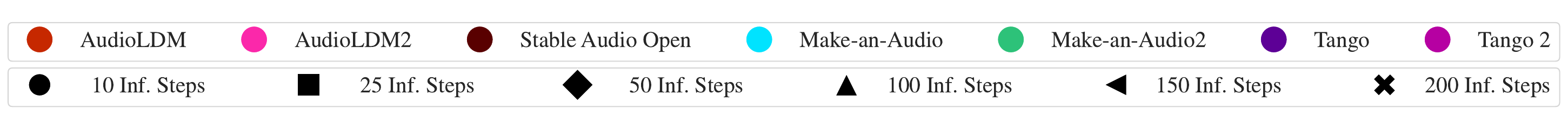}

  \caption{Pareto frontier for TTA models. The legend shows the model by color, and the point shape indicates the number of inference steps.}
  \label{fig:all_results}
\end{figure*}

\section{PARETO FRONTIER ANALYSIS}
\label{subsec:pareto}

While the previous two experiments provide interesting insights into the energy consumption of TTA models under different technical settings, real-world applications place greater emphasis on the quality of the generated sound and how well it matches the input prompt. 
These factors often shape the design of TTA generative systems. As such, it becomes essential to consider the trade-offs between audio quality, prompt alignment, and energy consumption, with the goal of achieving high-quality results while minimizing environmental impact. To understand how TTA models perform in practical scenarios, we apply Pareto optimization~\cite{douwes2023quality} across various inference configurations. Section~\ref{subsec:pa} discusses the background, methodological framework, datasets, and audio quality metrics used in this analysis.

We selected a subset of 300 audio samples from AudioCaps-test and Clotho-eval, to maintain a practical balance between generation size and energy consumption. Using the captions of the selected subsets, we generated dataset subsets with all TTA models, measured their energy consumption, calculated the FAD and CLAP metrics, and identified the corresponding Pareto frontier. All audio files were trimmed to 10 seconds for fair comparison. To validate the reliability of our findings, we conducted a sanity check by generating full versions of both datasets (AudioCaps and Clotho), using the original audio durations. For the sanity check, we used AudioLDM and Stable Audio Open, chosen as representative models due to their energy consumption. The sanity check confirmed that the trends observed with the subsets were consistent with those from the full datasets, supporting the decision to proceed with the smaller-scale evaluation.

Figure~\ref{fig:all_results} presents the Pareto frontier analysis of the TTA models, illustrating the trade-off between energy consumption (in kWh) during inference, measured by CodeCarbon, and audio quality, assessed using FAD and CLAP scores on the generated dataset. The plot uses a logarithmic scale. Each data point represents a unique model configuration, with marker shapes indicating the number of inference steps and colors denoting the model used. The models on the optimal Pareto frontier are highlighted with a red contour.

AudioLDM, AudioLDM2, Make-an-Audio, and Make-an-Audio-2 show similar average performance across both metrics and datasets, consistently performing in line with each other when compared to other models. Among these, AudioLDM stands out as the most energy-efficient, consistently appearing on the Pareto frontier with the lowest total energy consumption. In contrast, the more energy-intensive Tango and Tango2 models achieve high CLAP scores across both datasets, though their FAD performance tends to be more variable. Specifically, Tango is featured on the Pareto frontier for AudioCaps in terms of FAD and for both datasets in terms of CLAP, while Stable Audio Open achieves the lowest FAD score on Clotho. These performance differences are likely due to variations in the training data used. Stable Audio Open was trained on a custom dataset that combined Freesound~\cite{fonseca2017freesound} and the Free Music Archive~\cite{evans2025stable}, while Tango and Tango2 were trained on a more heterogeneous dataset, including AudioSet~\cite{gemmeke2017audio} and AudioCaps~\cite{kim2019audiocaps}. This diversity in training data probably contributes to their stronger performance on AudioCaps. Additionally, Tango2 benefits from further fine-tuning on the custom Audio-Alpaca dataset. These differences in training data and fine-tuning strategies might explain the observed performance variations: Stable Audio Open is better suited for capturing the characteristics of Clotho, while the Tango models excel on AudioCaps, reflecting the influence of their distinct training sources. 

Another important factor shaping the Pareto frontier is the number of inference steps used in TTA models. In our analysis, most Pareto-optimal points correspond to lower diffusion step counts, typically between $10$ and $50$. This indicates that increasing the number of diffusion steps beyond a certain point fails to yield improvements in FAD or CLAP scores, while energy cost continues to grow, rendering the additional steps inefficient and unproductive. This suggests that aggressive step scheduling or early stopping strategies could significantly reduce energy consumption without major quality loss. In fact, configurations with $150$ or $200$ diffusion steps never appear on the Pareto frontier, and the $100$-steps setting appears once.

\section{Discussion}
\label{sec:disc}

This section reflects on the findings of our experiments, highlighting their practical implications for energy-aware use of TTA models. 

The first two experiments clearly show that increasing the number of inference steps leads to higher energy consumption in TTA models. Moreover, when multiple outputs are needed for the same prompt, generating them using an optimal batch size is more efficient than producing them one at a time. The Pareto frontier analysis highlights that model selection should be guided by the specific needs of the application. Depending on the use case and audio characteristics, a model with lower energy consumption may be preferred over one with more inference steps, as excessive steps can increase energy use without clear quality gains (Fig.~\ref{fig:all_results}). However, sometimes a smaller model with more steps may offer a better trade-off than a larger model with fewer steps or vice versa (see Tango2 vs Make-an-audio2 in Fig.~\ref{fig:all_results}a and Fig.~\ref{fig:all_results}b). When higher quality is necessary, switching to a more advanced model may be necessary, but energy efficiency should remain a key consideration. The distribution of results shows that no single model excels in all scenarios. Instead, different models perform better depending on the optimization goal, whether it is maximizing quality or efficiency, underscoring the value of multi-objective analysis. Application-specific constraints, such as real-time response or offline processing, determine which region of the Pareto frontier is most relevant. This makes context-aware model deployment essential, particularly in edge or interactive settings. In particular, the performance gaps between configurations using $50$, $100$, and $200$ inference steps suggest that tuning the step count could yield better quality-efficiency trade-offs. Exploring this further is a promising direction for future work. Another important observation is the variation in model performance on FAD across datasets. This suggests that model effectiveness depends on both the characteristics of the target audio and the data it was trained on. As a result, the trade-off between perceptual quality and energy use can shift with the data domain. These findings reinforce the need to choose models that match the specific requirements and constraints of each generation.

\section{Conclusions}
\label{sec:conclusions}
This paper presents a first analysis of the energy consumption of diffusion-based TTA generative models during inference. Our study first analyzes how the number of inference steps and batch size affect energy consumption. To provide actionable insights, we then 
identify Pareto-optimal configurations using FAD and CLAP score. Our findings show that model energy consumption is influenced by multiple interacting factors and should be an integral consideration in the design of generative architectures. These results also highlight the need for deeper, component-level analyses of model architectures and motivate extending this work to other generative paradigms, such as auto-regressive models. While this paper raises open questions, we see it as a part of our contribution. Energy consumption in generative audio models is a persistent issue likely to grow, making early discussions and investigations crucial.

\clearpage
\bibliographystyle{IEEEtran}
\bibliography{refs25}







\end{document}